\documentclass[preprint,showpacs,preprintnumbers,amsmath,amssymb]{revtex4}

\usepackage{epsfig}
\usepackage{graphicx}
\usepackage{dcolumn}
\usepackage{bm}
\usepackage{color}

\let\jnfont=\rm
\def\NPB#1,{{\jnfont { \it Nucl.\ Phys.\ B, }}{\bf #1}}
\def\PLB#1,{{\jnfont { \it Phys.\ Lett.\ B, }}{\bf #1}}
\def\EPJC#1,{{\jnfont {\it Eur.\ Phys.\ Jour.\ C, }}{\bf #1}}
\def\EPL#1,{{\jnfont { \it EPL,\ }}{\bf #1}}
\def\PRD#1,{{\jnfont {\it Phys.\ Rev.\ D, }}{\bf #1}}
\def\PRL#1,{{\jnfont { \it Phys.\ Rev.\ Lett,\ }}{\bf #1}}
\def\MPLA#1,{{\jnfont{ \it  Mod.\ Phys.\ Lett,\ A, }}{\bf #1}}
\def\JPG#1,{{\jnfont {\it J.\ Phys.\ G,}}{\bf #1},}
\def\CTP#1,{{\jnfont{ \it Commun.\ Theor.\ Phys,\ }}{\bf #1}}
\def\CPL#1,{{\jnfont{ \it Chin.\ Phys.\ Lett,\ }}{\bf #1}}
\def\JHEP#1,{{\jnfont {\it JHEP, \ }}{\bf #1}}
\def\NPPS#1,{{\jnfont {\it Nucl.\ Phys.\ Proc.\ Suppl,\ }}{\bf #1}}

\begin{document}

\title{\ \\[10mm]
Associated production of the $Z$ boson with a pair of top quarks in
the left-right twin Higgs model }

\author{ \footnotesize Jinzhong Han$^{1}$}\email{hanjinzhongxx@gmail.com}
\author{\footnotesize Bingfang Yang $^{2,3}$}\email{yangbingfang@gmail.com}
\author{ \footnotesize Xiantu Zhang$^{1}$}

\affiliation{\footnotesize $^1$School of Physics and
Electromechnical  Engineering, Zhoukou Normal University, Henan,
466001, China\\ $^2$ Basic Teaching Department, Jiaozuo University,
Jiaozuo 454000, China \\ $^3$ School of Materials Science and
Engineering, Henan Polytechnic University, Jiaozuo 454000, China
   \vspace*{1.5cm}  }

\begin{abstract}
In the context of the left-right twin Higgs (LRTH) model, we first
examine the effects on the $Zt\bar t$ production at the ILC and LHC.
Our results show that the cross-sections can be significantly
deviated from the standard model predictions and thus provide a good
probe for the LRTH model. We also estimate the new production
channel, $Zt\bar T $ or $Z\bar t T $ production, at the LHC.
Compared with $Zt \bar t$ production, we find that the $Z t \bar T$
production can have a sizable production rate when the scale $f$ is
not too high. Considering the dominant decay mode $T\rightarrow
\phi^{+}b\rightarrow tb\bar b$, we find that $Z t\bar T $ final
state has less background than $Zt\bar t$ production and may likely
be observable at the LHC.

\end{abstract}


\maketitle

\section{Introduction}
The direct evidence for the top quark was presented in 1995 by the
CDF and D0 collaborations \cite{Tevatrontop}. From that time on, top
quark physics has always been one of the central physical topics at
the Large Hadron Collider (LHC) and the future International Linear
Collider (ILC). The top quark is the heaviest particle in the
standard model (SM), so it is widely speculated that the properties
of the top quark are sensitive to new physics. Deviations of
experimental measurements from the SM predictions would indicate new
non-standard top production or decay mechanisms. One of particular
interest is the large top quark forward-backward asymmetry observed
at the Tevatron may imply the new physics in the top quark sector
\cite{afb1}.

The probe of the couplings between the top quark and gauge bosons,
such as $\gamma t\bar{t},Zt\bar{t}$ and $Wtb$, is another way to
discover new physics. Because of the small cross-section,
$Zt\bar{t}$ coupling is not observable at the Tevatron. On the
contrary, these couplings can be measured precisely at the LHC
\cite{zttLHC}and the ILC \cite{zttILC}. And many relevant works
focusing on $pp\rightarrow Zt\bar{t}$ at the LHC \cite{pp-ztt} and
$e^{+}e^{-}(\gamma\gamma)\rightarrow Zt\bar{t}$ at the ILC
\cite{ee-ztt} in the SM and beyond have been done. Recently, the CMS
Collaboration \cite{zttCMS}and the ATLAS Collaboration
\cite{zttATLAS} have, respectively, published the first set of
results using the $\sqrt{s}=7$ TeV $pp$ collision data by the
trilepton channel, in which the $Z$ boson decays to a pair of
leptons and one of the $W$ bosons coming from $t\rightarrow Wb$
decays, gives rise to a lepton after decay. The measured values are
compatible within uncertainties with the next-to-leading order(NLO)
SM calculations.

To solve the little hierarchy problem of the SM, the left-right twin
Higgs (LRTH) model was proposed and regarded as an alternative
candidate for new physics \cite{LRTH-1,LRTH-2,LRTH-3}. The
phenomenology of the LRTH model has been widely discussed in Refs.
\cite{phenomenology-1,phenomenology-2}. In this model, a top partner
(denoted as $T$-quark) is contained. Due to the mixing between $t$
and $T$, the $Zt\bar{t}$ coupling is modified. Moreover, the
$T$-quark can contribute to the $ Zt\bar{t}$ production process
through its virtual effects. The precision measurements of $
Zt\bar{t}$ production at high energy colliders make it possible to
unravel the new physics effects or constrain the model parameters.
In addition, the new production channel $Zt\bar T $ or $Z\bar t T $
productions can be implemented at the LHC. This new production
channel has the different final states from the $Zt\bar{t}$
production due to the dominant decay $T\rightarrow
\phi^{+}b\rightarrow t\bar{b}b$. A search for this new effect will
provide a good probe to detect the LRTH model.

This paper is organized as follows. In Sec.II we briefly review the
LRTH model related to our calculations. We study the $Zt\bar{t}$
production at the ILC in Sec.III and the $Zt\bar{t}$ production at
the LHC in Sec.IV, respectively. In Sec.V we study the new
production channel $Zt \bar T$ or $ZT\bar t$ at the LHC. Finally, we
give our conclusions in Sec.VI.

\section{A BRIEF REVIEW OF THE LRTH MODEL}
Here we will briefly review the ingredients which are relevant to
our calculations, and a detailed description of the LRTH model can
be found in Ref. \cite{phenomenology-1}.

The LRTH model is based on the global $U(4)_1\times U(4)_2$ symmetry
with a locally gauged subgroup $SU(2)_L\times U(2)_R\times
U(1)_{B-L}$. Under the global symmetry, two Higgs fields,
$H=(H_L,H_R)$ and  $\hat{H} =({\hat{H}_L},\hat{H}_R$), are
introduced and each transforms as (4,1) and (1,4), respectively.
${H}_{L,R} ({\hat{H}_{L,R}})$ are two component objects which are
charged under $SU(2)_L$ and $SU(2)_R$, respectively. The global
$U(4)_1[U(4)_2$] symmetry is spontaneously broken down to its
subgroup $U(3) [U(3)]$ with non-zero vacuum expectation values
(VEVs) as $<H>$ =$(0,0,0,f)$ and $H$ =$(0,0,0,\hat{f})$. Each
spontaneously symmetry breaking results in seven Nambu-Goldstone
bosons. Three Goldstone bosons are eaten by the massive heavy gauge
bosons $W^{\pm}_H$ and $Z_H$, while the remaining Goldstone bosons
contain three physical Higgs $\phi^{0}$ and $\phi^{\pm}$. The mass
of the heavy gauge bosons can be expressed as:
\begin{eqnarray}
M_{W_{H}}^{2}&=& \frac{1}{2}g^{2}(\hat{f}^{2}+f^{2}\cos^{2}x),\\
M_{Z_{H}}^{2}&=&
\frac{g^{2}+g'^{2}}{g^{2}}(M_{W}^{2}+M_{W_{H}}^{2})-M_{Z}^{2},
\end{eqnarray}
where $g=e/{S_W}$, $g'=e/{\sqrt{\cos2\theta_W}}$,
$S_W=\sin\theta_W$, $\theta_{W}$ is the Weinberg angle.
$x=v/(\sqrt{2}f)$, and $v$ is the electroweak scale, the values of
$f$ and $\hat{f}$ will be bounded by electroweak precision
measurements. In addition, $f$ and $\hat{f}$ are interconnected once
we set $v$ =246 GeV.

The mass of the light SM-like top quark and its partner  heavy top
quark $T$ are
\begin{eqnarray}
m_{t}^{2} = \frac{1}{2}(M^{2}+y^{2}f^{2}-N_{t}),\hspace{0.5cm}
M_{T}^{2} = \frac{1}{2}(M^{2}+y^{2}f^{2}+N_{t}),
\end{eqnarray}
where $N_{t} = \sqrt{(y^{2}f^{2}+M^{2})^{2}-y^{4}f^{4}\sin^{2}2x}$.
Provided $M_{T} \leq f$ and the parameter $y$ is of order one, the
top Yukawa coupling will also be of order one. The mass parameter
$M$ is essential to the mixing between the SM top quark and its
partner $T$. At the leading order of $1/f$, the mixing angles can be
written as:
\begin{eqnarray}
S_L=\sin\alpha_L\simeq\frac{M}{M_T}\sin{x},~~~~~~S_R=\sin\alpha_R\simeq\frac{M}{M_T}(1+\sin^2{x}),
\end{eqnarray}
The couplings expression forms which are related to our calculations
are given as follows \cite{phenomenology-1}:
\begin{eqnarray}
&&g^{Zt\bar{T}}_L=\frac{eC_{L}S_{L}}{2C_{W}S_{W}},~~~~~~~~~~~~~~{{g^{Zt\bar{T}}_R=\frac{ef^2x^2S_WC_{R}S_{R}}{2\hat {f}^2C^3_{W}}}}; \\
&&{{g^{Zt\bar{t}}_L=\frac{e({3C^2_{{L}}-4S^2_W)}}{6C_WS_W},~~~~~~~g^{Zt\bar{t}}_R=-\frac{{2eS^2_W}}{3C_WS_W}}};\\
&&g^{Ze^{+}e^{-}}_L=\frac{e({-\frac{1}{2}+S^2_W)}}{S_WC_W},~~~~~~g^{Ze^{+}e^{-}}_R=\frac{eS_W}{C_W};\\
&&g^{Z_Ht\bar{T}}_L=\frac{eC_{L}S_{L}S_W}{2C_{W}\sqrt{\cos2\theta_W}},~~~~~g^{Z_Ht\bar{T}}_R=-\frac{eC_{R}S_{R}C_W}{2S_{W}\sqrt{\cos2\theta_W}};\\
&&{{g^{Z_He^{+}e^{-}}_L=\frac{eS_W}{2C_{W}\sqrt{\cos2\theta_W}}}},~~g^{Z_He^{+}e^{-}}_R=\frac{e(1-3\cos2\theta_W)}{4S_WC_{W}\sqrt{\cos2\theta_W}}; \\
&&{{V_{hZ_{\mu}Z_{\nu}}=\frac{em_Wg_{\mu\nu}}{S_WC^2_{W}},~~~~~~~~~~~V_{hZ_{\mu}Z_{H_\nu}}=\frac{e^2fxg_{\mu\nu}}{\sqrt{2}C^2_{W}\sqrt{\cos2\theta_W}}}};\\
&&V_{ht\bar{t}}=-\frac{em_tC_LC_R}{2m_WS_{W}};
\end{eqnarray}
where $C^2_L=(1-S^2_L)$, $C^2_R=(1-S^2_R)$.
\section{Production of $Z t \bar t$  at the ILC}

In this section, we study the process $e^{+}e^{-}\to Zt\bar t$ in
the LRTH model at the ILC. The relevant Feynman diagrams are shown
in fig.1. In comparison with the SM, we can see there are additional
diagrams mediated by the ${Z_H}$ gauge boson and the heavy $T$-quark
in the LRTH model.
\begin{figure}[tbh]
\begin{center}\epsfig{file=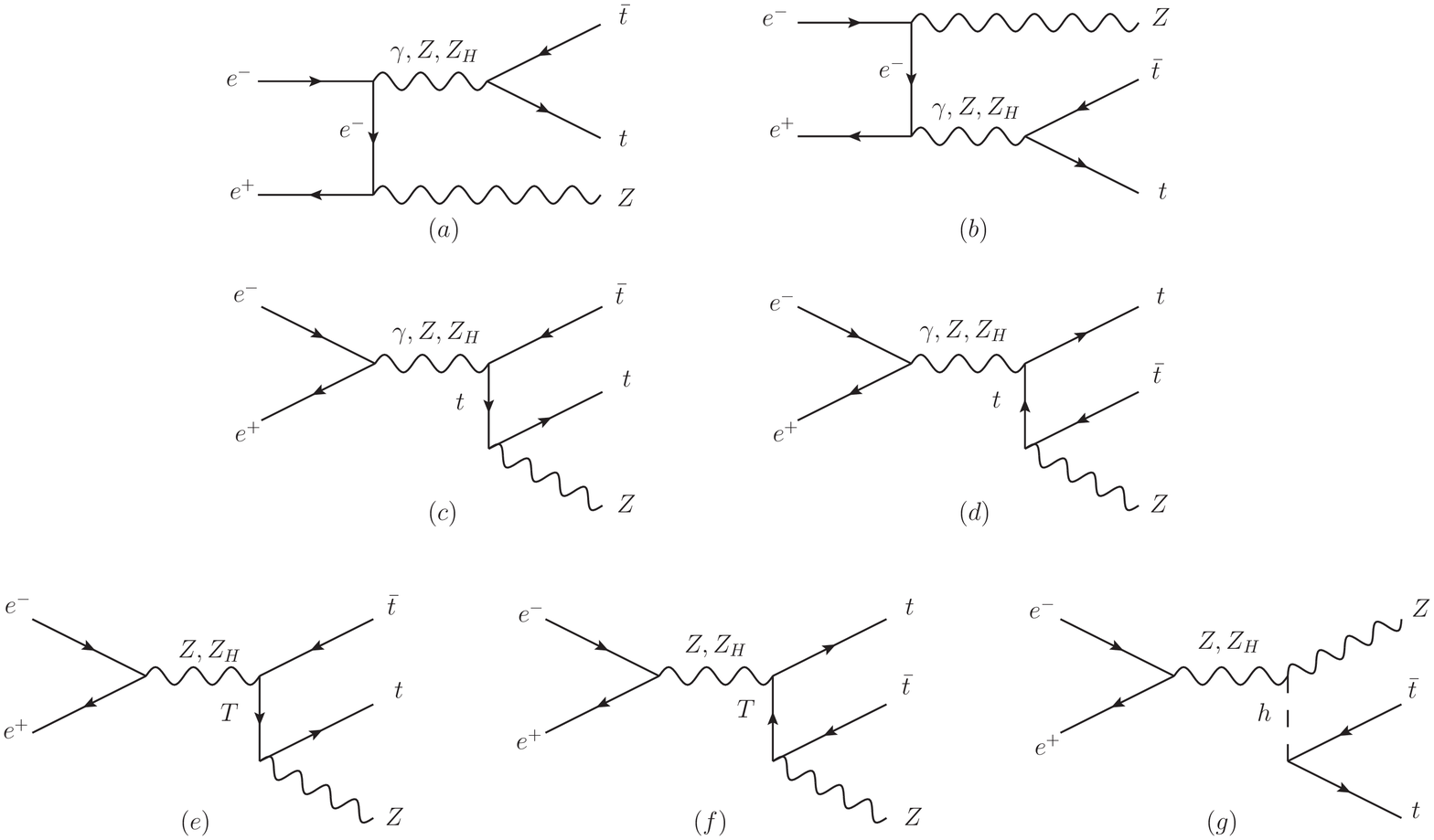,width=14cm}
\end{center}\vspace{-0.5cm}
\caption{Feynman diagrams for $e^{+}e^{-}\to Zt\bar t$ in the LRTH
model at the ILC.}
\end{figure}

In our numerical calculations, we take the SM parameters as follows
\cite{SM-paramet}:
\begin{eqnarray}
\alpha(m_{Z})=1/128.8, \sin^2\theta_{W}=0.231,~~~~~~~~~~~~\nonumber\\
m_{t}=172.4 ~{\rm GeV}, m_{h}=125 ~{\rm GeV}~ [15],
m_{Z}=91.2 ~{\rm GeV}.
\end{eqnarray}
In addition, there are some LRTH model parameters involved in the
amplitudes, they are $f$($\hat f$) and $M$. The parameter $\hat f$
can be determined by requiring that the SM Higgs boson obtains an
electroweak symmetry breaking VEV of 246 GeV. The top Yukawa
coupling constant $y$ can also be determined by fitting the
experimental value of the top quark mass $m_t$. Following Ref.
\cite{phenomenology-1}, we vary the scale $f$ in the range of $500$
GeV $\leq f \leq  $1500 GeV and take the mixing parameter $M$ = 100
GeV, 150 GeV, 200 GeV as an example.

\begin{figure}[htbp]
\scalebox{0.7}{\epsfig{file=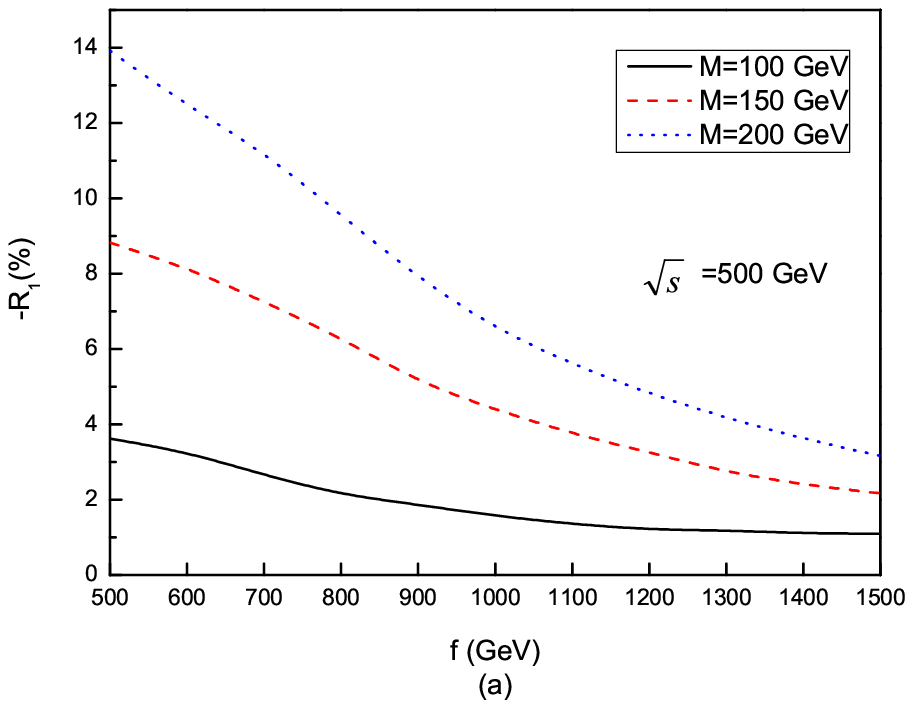}}
\scalebox{0.7}{\epsfig{file=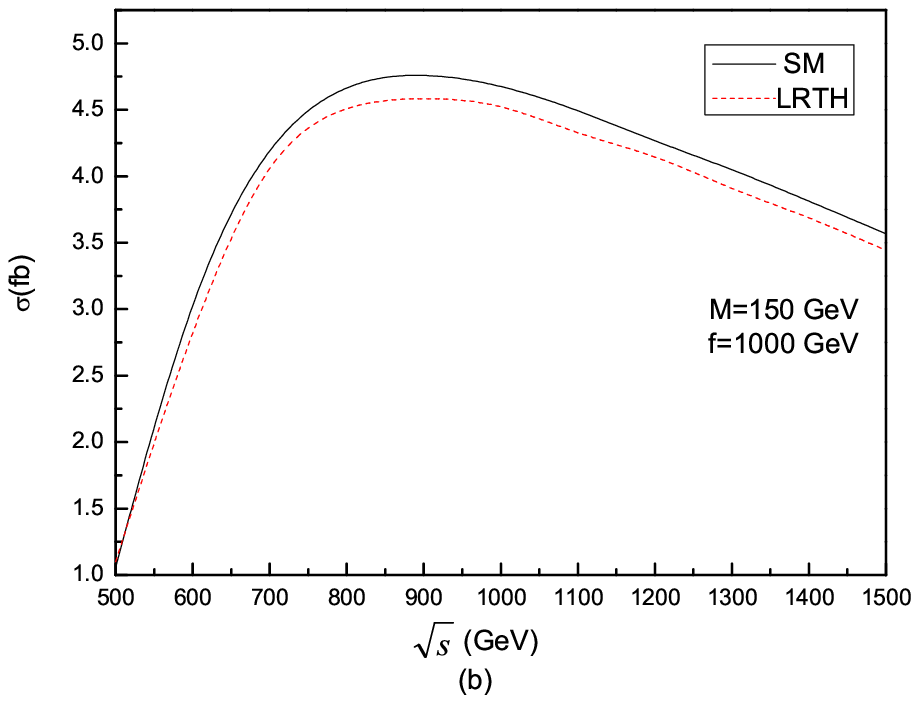}} \vspace{-0.5cm}\caption{
(a) The deviation from the SM prediction of the $Zt\bar t$
production cross section versus the scale $f$ and (b) the $Zt\bar t$
production cross section as functions of the center-of-mass energy
$\sqrt{s}$.}
\end{figure}

In fig.2(a), we show the deviation from the SM prediction of the
$Zt\bar t$ production cross-section
$R_{1}=(\sigma^{LRTH}-\sigma^{SM})/\sigma^{SM}$ as function of the
scale $f$ for the three values of the mixing parameter $M$ at the
ILC for $\sqrt{s}=500$ GeV. We can see that the deviation is
negative so that the LRTH contributions decrease the SM
cross-section. When the scale $f$ increases, the deviation from the
SM prediction $R_{1}$ become small, which indicates that the effects
of the LRTH model will decouple at the high scale $f$. The maximum
value of the deviation from the SM prediction $R_{1}$ can reach
$-14\%$ in the allowed parameter space. In fig.2(b), we show the
production cross section $\sigma$ as function of center-of-mass
energy $\sqrt{s}$ in the LRTH model and the SM for $f=1000$ GeV, $M$
= 150 GeV at the ILC, respectively. Since the process proceeds
mainly through the s-channel, we can see that the $t\bar t Z$
cross-sections first increase and then decrease with the increasing
values of $\sqrt{s}$.

According to the Ref.\cite{ILCttz}, if only one $t\bar tV(V=\gamma,
Z)$ coupling at a time is allowed to deviate from its SM value, a
linear $e^{+}e^{-}$ collider operating at $\sqrt{s} = 500$ GeV with
an integrated luminosity of $100\sim 200 ~{\rm fb}^{-1}$ would be
able to probe all $Zt\bar t$ couplings with a precision of $1\sim
5\%$.

\section{Production of $Z t \bar t$  at the LHC}
The production of $Z t\bar t$ at the LHC can proceed through $gg$
fusion or $q\bar{q}$ annihilation, the relevant Feynman diagrams are
shown in fig.3.

\begin{figure}[tbh]
\begin{center}\epsfig{file=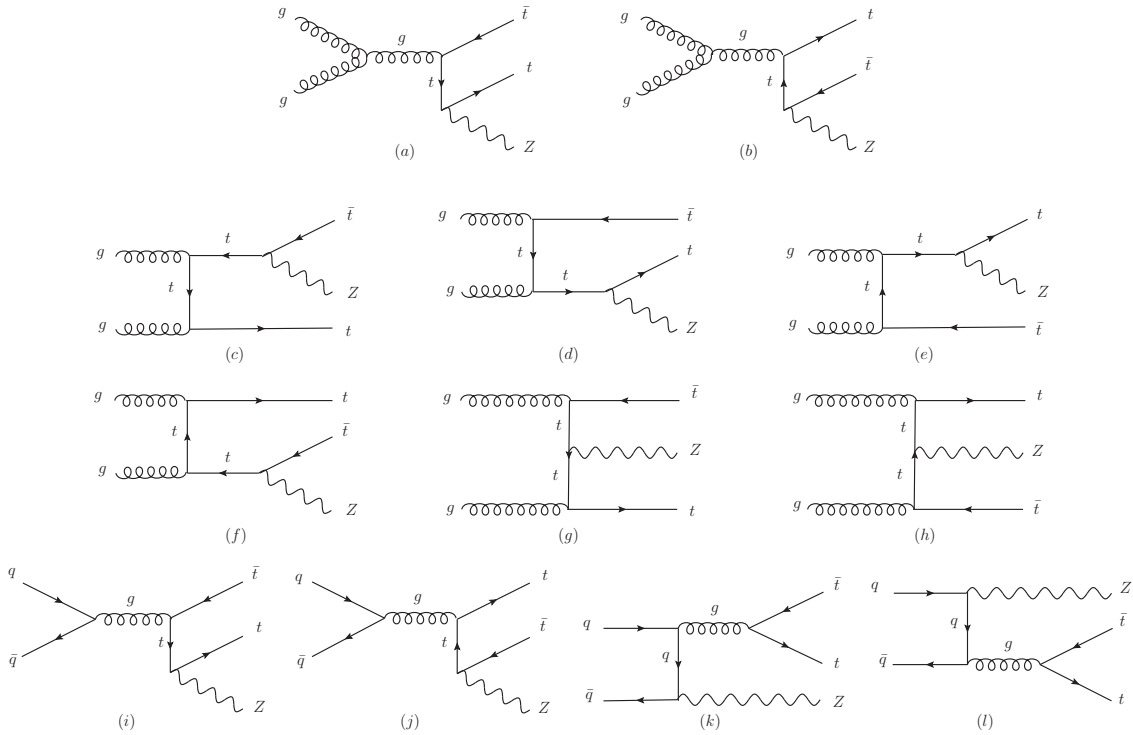,width=15.5cm}
\end{center}\vspace{-0.5cm}
\caption{Feynman diagrams for $Zt\bar t$ production in the LRTH
model at the LHC.}
\end{figure}

The relevant SM parameters are taken as follows
\begin{eqnarray}
m_{t}=175 ~{\rm GeV}, m_{Z}=91.2 ~{\rm GeV},~~~~~~~~\nonumber\\
\alpha(m_{Z})=1/128, \alpha_{s}=0.1172, \sin^2\theta_{W}=0.231.
\end{eqnarray}

For the relevant LRTH parameters, we also take $500~ {\rm GeV}\leq
f\leq 1500~ {\rm GeV}$ and $M$ = 100 GeV, 150 GeV, 200 GeV. In our
calculations, we used the CTEQ5M patron distribution functions
\cite{CTEQ5M}.

\begin{figure}[tbh]
\scalebox{0.7}{\epsfig{file=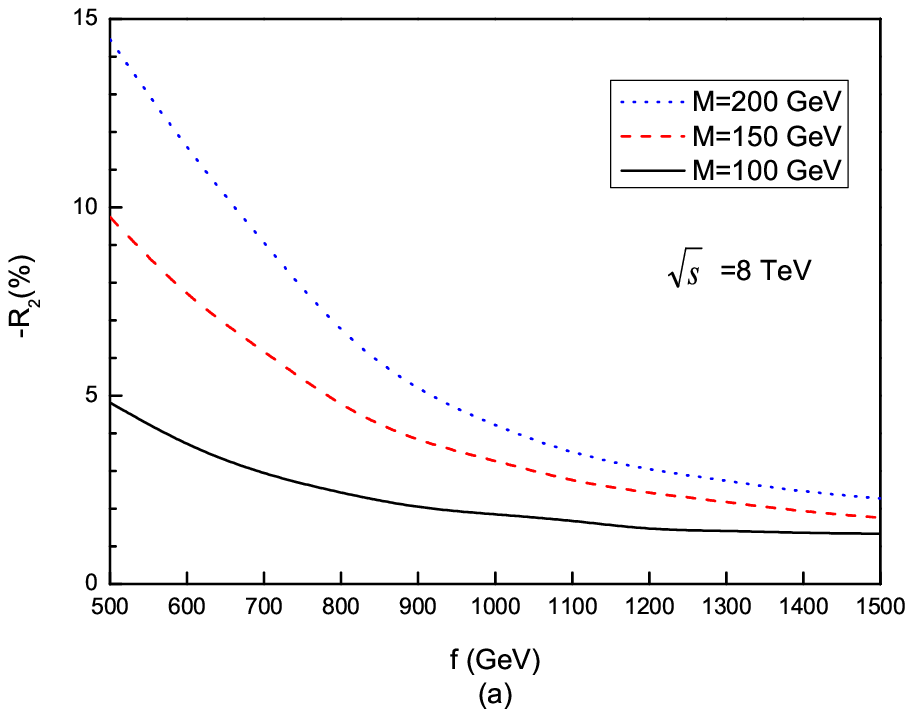}}
\scalebox{0.7}{\epsfig{file=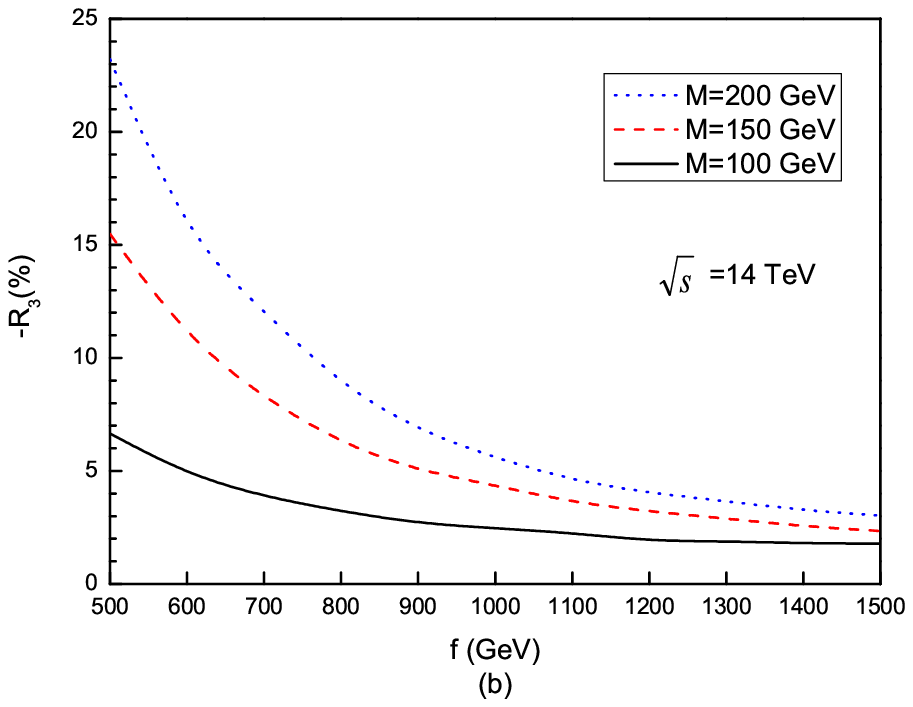}}
\vspace{-0.5cm}\caption{The deviation from the SM prediction of the
$Zt\bar t$ production cross section as functions ofthe scale $f$ for
$\sqrt{s}=8$ TeV(a) and $\sqrt{s}=14$ TeV(b), respectively.}
\end{figure}

In fig.4(a) and fig.4(b) we show the deviation from the SM
prediction $R_{2}(R_{3})=(\sigma^{LRTH}-\sigma^{SM})/\sigma^{SM}$ of
the $Zt\bar t$ production cross-section as a functions of the scale
$f$ for $\sqrt{s}=8$ TeV and $\sqrt{s}=14$ TeV, respectively. We can
see that the deviation from the SM prediction parameter $R_2$ can
reach $-14$\% and $R_3$ can reach $-23$\%. According to Ref.
\cite{ttzMeasurement}, the improvement is particularly pronounced
for the $Zt\bar t$ axial vector coupling which can be measured with
a precision of $3\sim 5\%$ at the luminosity-upgraded LHC
(3000~${\rm fb}^{-1}$). From these two figures we also can see that
the deviation from the SM prediction decrease the SM cross-section
in the allowed parameter space, which makes the observation of this
production channel even harder.

\section{Productions of $Z t \bar T$ and  $Z T \bar t$ at the LHC}
Like the  $Z t \bar t$ production, the new production channel $Z t
\bar T$ or $Z T \bar t$ can proceed through $gg$ fusion or
$q\bar{q}$ annihilation at the LHC, the relevant Feynman diagrams
are shown fig.5.
\begin{figure}[tbh]
\begin{center}\epsfig{file=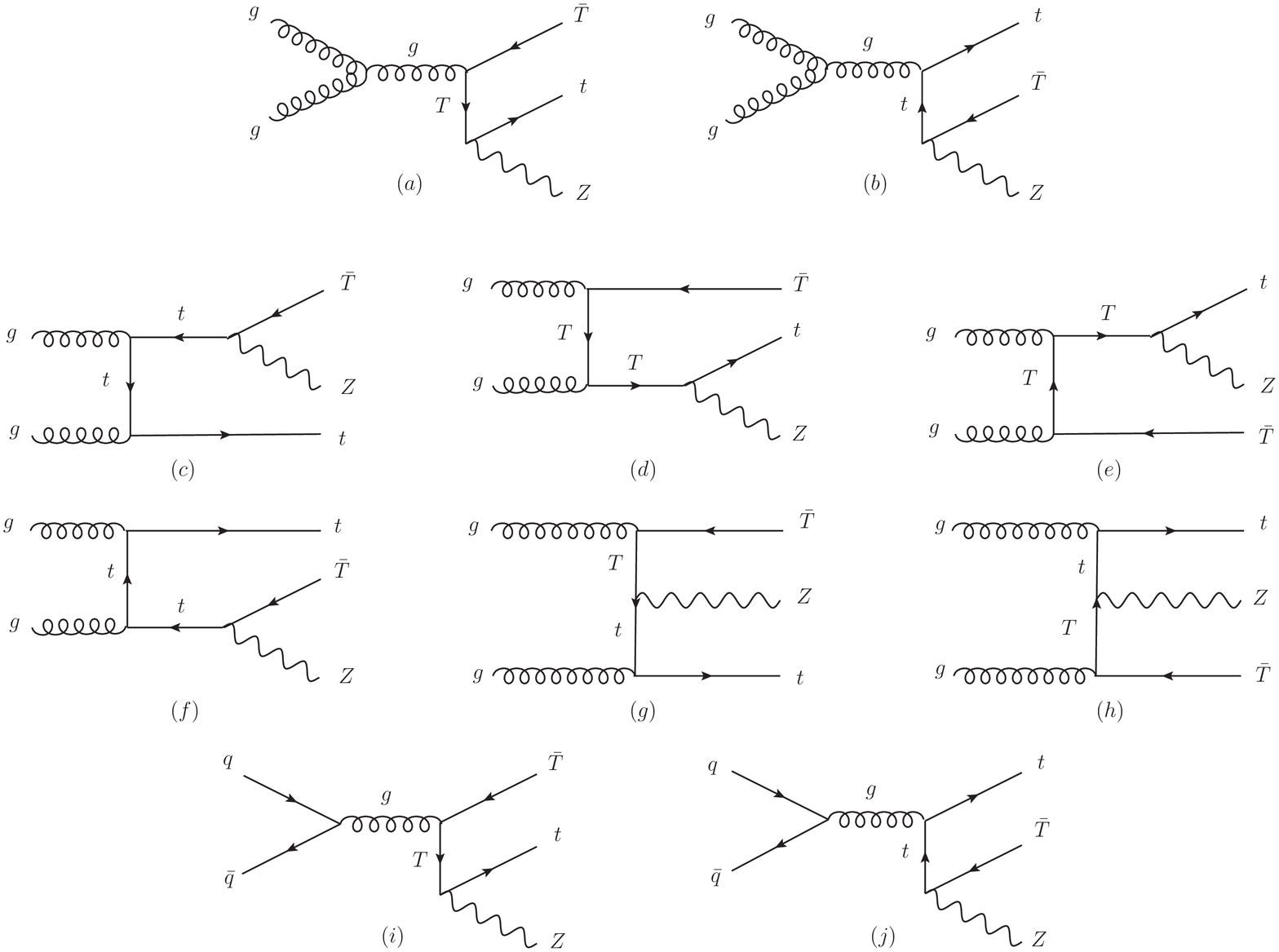,width=14cm}
\end{center}\vspace{-0.5cm}
\caption{Feynman diagrams for $Z t \bar T$ production in the LRTH
model at the LHC. }
\end{figure}

\begin{figure}[htb]
\scalebox{0.7}{\epsfig{file=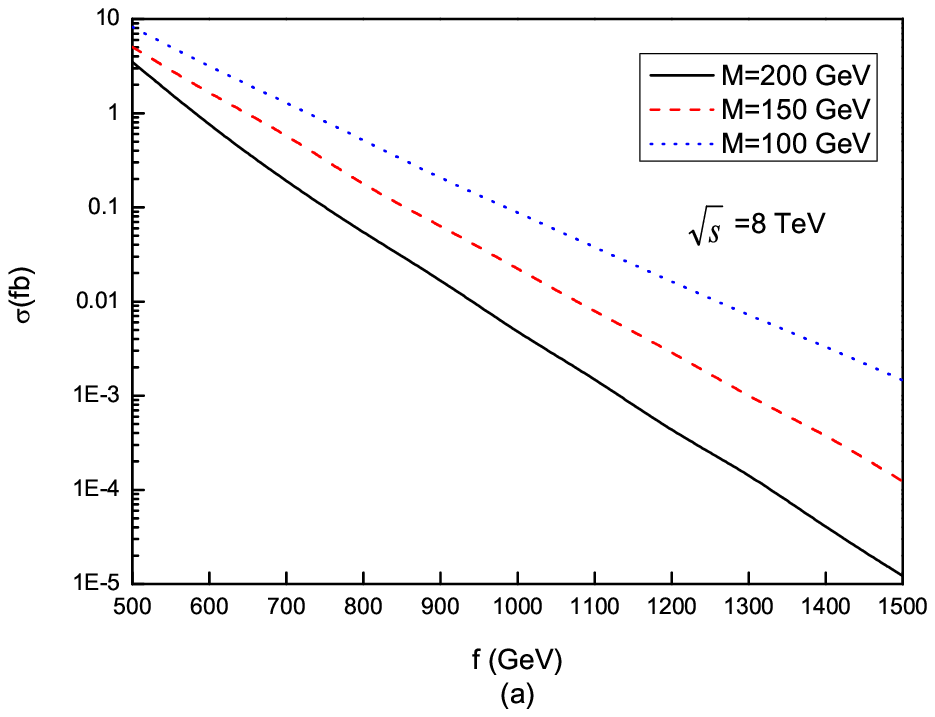}}
\scalebox{0.7}{\epsfig{file=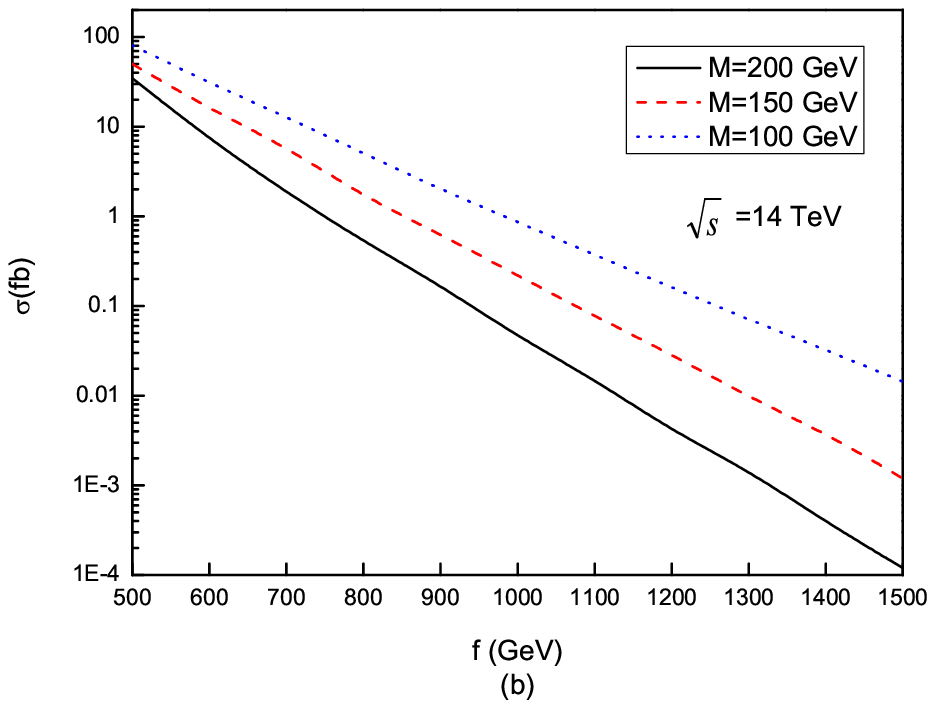}}
\vspace{-0.5cm}\caption{The ($Zt\bar{T}+ZT\bar{t}$) production cross
section as functions of the scale $f$ for $\sqrt{s}=8$ TeV (a) and
$\sqrt{s}=14$ TeV(b), respectively.}
\end{figure}
In fig.6 we plot the ($Zt\bar{T}+ZT\bar{t}$) production
cross-section $\sigma$ as a function of the scale $f$ and the three
values of the mixing parameter $M$ for $\sqrt{s}=8$ TeV and
$\sqrt{s}=14$ TeV, respectively. We can see that the production
cross-section $\sigma$ is very sensitive to the parameters $f$ and
$M$. On the other hand, the cross-section value decreases quickly as
the parameter $f$ increase. For $\sqrt{s}=14$ TeV, $M$ =150 GeV and
500 GeV $\leq f\leq $ 1000 GeV, the value of the total hadronic
cross-section is in the range $49.8{\rm fb}\sim 0.22{\rm fb}$. For
the anticipated integrated luminosity of 100~${\rm fb}^{-1}$ even
for a high integrated luminosity of 1000~${\rm fb}^{-1}$, when the
parameter $f$ is not too high the LHC will copiously produce the $Z
T\bar t$ events per year.

It has been shown that the branching ratio of $\phi^{+}\rightarrow
t\bar{b}$ is approximately equal to $100\%$ for the mixing parameter
$M>10$ GeV \cite{phenomenology-1}. Thus, the dominant decay mode
$T\rightarrow \phi^{+}b\rightarrow t\bar{b}b$ can make the processes
$pp\rightarrow ZT\bar t + Zt\bar T$ give rise to the
$t\bar{t}b\bar{b}b\bar{b}$ final state with $Z\rightarrow b\bar{b}$.
For this final state, the main backgrounds come from the SM
processes $pp\rightarrow t\bar{t}ZZ + X$ and $pp\rightarrow
t\bar{t}hh + X$ with $Z\rightarrow b\bar{b}$ and $h\rightarrow
b\bar{b}$, where the additional jets (light  quarks or gluons) may
be misidentified as $b$-quarks. The relevant studies
\cite{background} have found that the largest background
$t\bar{t}b\bar{b}jj$ can be suppressed by enhancing the ability to
tag $b$-jets. Furthermore, a systematic signal-to-background
analysis including the observability of the processes $pp\rightarrow
ZT\bar t + Zt\bar T$ would depend on Monte Carlo simulations, which
is beyond the scope of our discussion.

\section{Conclusions}
In this paper, we studied the top quark pair production associated
with a $Z$ boson in the LRTH model at the ILC and the LHC. For the
production of $Z t\bar{t}$ at the ILC, we found that the deviation
from the SM prediction of the cross-section can reach over $10\%$ in
magnitude and this effect should be observable \cite{ILCtop}. For
the production of $Z t\bar{t}$ at the LHC, we found that the
deviation from the SM prediction of the cross-section can reach over
$5\%$ in magnitude when the scale $f<800$ GeV. For the new
production channel of $Zt\bar{T}$ or $ZT\bar{t}$, we found  that the
$Z t \bar T$ production can have a sizable production rate when the
scale $f$ is not too high. Considering the dominant decay mode
$T\rightarrow \phi^{+}b\rightarrow t\bar{b}b$ with $Z\rightarrow
b\bar{b}$, the production of $Zt\bar{T}$ or $ZT\bar{t}$ may have the
less background than the $Zt\bar{t}$ production, and thus this new
channel may likely be observable at the LHC.

\vspace{0.5cm} \noindent{\bf Acknowledgments:}

We appreciate the helpful suggestions from Jinmin Yang and thank
Yaobei Liu for useful discussions. This work is supported by the
National Natural Science Foundation of China (NNSFC) under grant
Nos. 11347140,11305049, and by the Education Department of Henan
under Grant No. 13A140113.

\end{document}